\documentclass[tightenlines,twoside,secnumarabic,
               onecolumn,floatfix,nofootinbib,showpacs,11pt]{revtex4}

\usepackage[english]{babel}
\usepackage{amsmath,amssymb}
\usepackage{graphicx}
\usepackage[sort&compress]{natbib}
\usepackage{slashed}            
\usepackage{hyperref}
\allowdisplaybreaks

\bibpunct{[}{]}{,}{n}{}{,}

\newcommand{\bra}[1]{\ensuremath{\langle #1 |}}   
\newcommand{\ket}[1]{\ensuremath{| #1 \rangle}}   
\newcommand{\ev}[1]{\ensuremath{\left\langle #1 %
                     \right\rangle}} 
\def\lesssim{\mathrel{\rlap{\lower4pt\hbox{\hskip1pt$\sim$}}
    \raise1pt\hbox{$<$}}}                
\def\gsim{\mathrel{\rlap{\lower4pt\hbox{\hskip1pt$\sim$}}
    \raise1pt\hbox{$>$}}}                

\newcommand{\gev}{{\rm GeV}}

\def\be{\begin{equation}}
\def\ee{\end{equation}}
\def\bea{\begin{eqnarray}}
\def\eea{\end{eqnarray}}

\newcommand{\met}{\slashed {E}_{T}}
\usepackage{xcolor}
\definecolor{red}{rgb}{1.0, 0, 0}

\begin{document}

\title{Missing Energy Signatures of Dark Matter at the LHC}
\author{Patrick J. Fox$^1$}          \email[Email: ]{pjfox@fnal.gov}
\author{Roni Harnik$^1$}             \email[Email: ]{roni@fnal.gov}
\author{Joachim Kopp$^1$}            \email[Email: ]{jkopp@fnal.gov}
\author{Yuhsin Tsai$^{2,1}$}         \email[Email: ]{yt237@cornell.edu}
\affiliation{$^1$ Theoretical Physics Department \\
                  \mbox{Fermilab, P.O.~Box 500, Batavia, IL 60510, USA} \vspace{0.2cm}\\
             $^2$ Institute for High Energy Phenomenology \\
                  Newman Laboratory of Elementary Particle Physics \\
                  Cornell University, Ithaca, NY 14853, USA}
\date{\today} 
\pacs{}

\begin{abstract}
  We use ATLAS and CMS searches in the mono-jet + missing energy and mono-photon +
  missing energy final state to set limits on the couplings of dark matter to
  quarks and gluons. Working in an effective field theory framework we compare several existing mono-jet analyses and find that searches with high $p_T$ cuts are more sensitive to dark matter. We 
  constrain the suppression scale of the effective dark matter--Standard Model
  interactions, and convert these limits into bounds on the cross sections
  relevant to direct and indirect detection. We find that, for certain types of
  operators, in particular spin-independent dark matter--gluon couplings and
  spin-dependent dark matter--quark couplings, LHC constraints from the
  mono-jet channel are competitive with, or superior to, limits from direct
  searches up to dark matter masses of order 1~TeV. Comparing to indirect
  searches, we exclude, at 90\% C.L., dark matter annihilating to quarks with
  the annihilation cross section of a thermal relic for masses below $\sim$
  15--70~GeV, depending on the Lorentz
  structure of the effective couplings. Mono-photon limits are somewhat weaker
  than mono-jet bounds, but still provide an important cross check in the
  case of a discovery in mono-jets. We also discuss the possibility that
  dark matter--Standard Model interactions at LHC energies cannot be described
  by effective operators, in which case we find that constraints can become
  either significantly stronger, or considerably weaker, depending on the mass and width of
  the intermediate particle. We also discuss the
  special case of dark matter coupling to the Higgs boson, and we show that searches for invisible Higgs decays would provide superior
  sensitivity, particularly for a light Higgs mass and light dark matter.
\end{abstract}

\begin{flushright}
  FERMILAB-PUB-11-478-T
\end{flushright}

\maketitle

\section{Introduction}

With the LHC physics program underway at full steam, the search for a dark
matter (DM) candidate in high energy collisions is gaining momentum. Missing energy
signatures are an integral part of many discovery channels for new physics at the
LHC, and if a deviation from Standard Model (SM) predictions should be found in any
of these channels, it could provide important evidence for the existence of
new particles that are stable (or at least long-lived) and neutral, thus fulfilling
two important requirements for being the dark matter in the universe.

In this paper, we will consider some of the more model-independent signatures of
dark matter at the LHC: events with a large amount of missing energy ($\met$) and a single jet or
a single photon, as well as missing energy signals associated with invisible decays of the Higgs boson.
Where available, we will use existing
LHC data to set limits on the dark matter--quark and dark matter--gluon couplings in an effective
field theory framework, and we will demonstrate the complementarity of these
limits to those obtained from direct and indirect dark matter searches.
We will also compare several mono-jet analyses that have been carried out by ATLAS and CMS, and we will outline a strategy for discovering dark matter or improving bounds in the future.

Dark matter searches using mono-jet signatures have been discussed previously
in the context of both Tevatron and LHC searches~\cite{Cao:2009uw,Agrawal:2010fh,Goodman:2010yf,
Bai:2010hh, Goodman:2010ku, Kopp:2011eu, Rajaraman:2011wf}, and have been shown
to be very competitive with direct searches, especially at low dark matter mass
and for dark matter with spin-dependent interactions. In a related work, SSC constraints on missing energy signatures due to quark and lepton compositeness have been discussed in~\cite{Chivukula:1987wf}. The mono-photon channel
has so far mostly been considered as a search channel at lepton
colliders~\cite{Birkedal:2004xn, Borodatchenkova:2005ct, Fox:2011fx}, but
sensitivity studies exist also for the LHC~\cite{Gershtein:2008bf,Wang:2011sx}, and they suggest
that mono-photons can provide very good sensitivity to dark matter production
at hadron colliders.  Combined analyses of Tevatron mono-jet searches and LEP
mono-photon searches have been presented in~\cite{Mambrini:2011pw,
Fortin:2011hv}.  The mono-photon channel suffers from different systematic
uncertainties than the mono-jet channel, and probes a different set of DM--SM
couplings, it can thus provide an important confirmation in case a signal is
observed in mono-jets. 

The outline of this paper is as follows: After introducing the effective field
theory formalism of dark matter interactions in section~\ref{sec:formalism}, we
will first discuss the mono-jet channel in section~\ref{sec:monojet}. We will
describe our analysis procedure and then apply it to ATLAS and CMS data in
order to set limits on the effective dark matter couplings to quarks and
gluons.  We also re-interpret these limits as bounds on the scattering and
annihilation cross sections measured at direct and indirect detection
experiments.  We then go on, in section~\ref{sec:lightmediators}, to discuss
how our limits are modified in models in which dark matter interactions are
mediated by a light $\lesssim \mathcal{O}(\text{few TeV})$ particle, so that
the effective field theory formalism is not applicable.  In
section~\ref{sec:monophoton}, we will perform an analysis similar to that from
section~\ref{sec:monojet} in the mono-photon channel.   A special example of
dark matter coupling through a light mediator is DM interacting through the
Standard Model Higgs boson, and we will argue in section~\ref{sec:invhiggs}
that in this case, invisible Higgs decay channels provide the best
sensitivity. We will summarize and conclude in section~\ref{sec:conclusions}.

\section{An Effective Theory for dark matter interactions}
\label{sec:formalism}

If interactions between dark matter and Standard Model particles involve
very heavy ($\gtrsim \text{few TeV}$) mediator particles---an assumption we are
going to make in most of this paper---we can describe them in the
framework of effective field theory. (We will investigate how departing from the effective
field theory framework changes our results in sections~\ref{sec:lightmediators} as well as \ref{sec:invhiggs}.)
Since our goal is not to do a full survey of all possible effective operators,
but rather to illustrate a wide variety of phenomenologically distinct cases,
we will assume the dark matter to be a Dirac fermion $\chi$ and consider the
following effective operators\footnote{Other recent studies that
have used a similar formalism to describe dark matter interactions
include~\cite{Harnik:2008uu,Cao:2009uv,Cao:2009uw,Agrawal:2010fh,Goodman:2010yf,
Bai:2010hh,Goodman:2010ku,Fan:2010gt,Goodman:2010qn,Fox:2011fx,Kamenik:2011nb,Rajaraman:2011wf,Cheung:2010kx,Cheung:2011uq,Cheung:2010vn}.}
\begin{align}
  \mathcal{O}_V &= \frac{(\bar\chi\gamma_\mu\chi)(\bar q \gamma^\mu q)}{\Lambda^2} \,,
    & \text{(vector, $s$-channel)} \label{eq:OV} \\
  \mathcal{O}_A &= \frac{(\bar\chi\gamma_\mu\gamma_5\chi)(\bar q \gamma^\mu\gamma_5 q)}{\Lambda^2} \,,
    & \text{(axial vector, $s$-channel)} \label{eq:OA} \\
  \mathcal{O}_t &= \frac{(\bar \chi P_R q)(\bar q P_L \chi)}{\Lambda^2} + (L\leftrightarrow R)\,,
    & \text{(scalar, $t$-channel)} \label{eq:OSt} \\
   \mathcal{O}_g &= \alpha_s \frac{(\bar\chi \chi)\,(G^a_{\mu\nu}G^{a\mu\nu})}{\Lambda^3} \,.
    & \text{(scalar, $s$-channel)} \label{eq:Og}
\end{align}
In these expressions, $\chi$ is the dark matter field, $q$ is a Standard
Model quark field, $G^a_{\mu\nu}$ is the gluon field strength tensor, and $P_{R(L)} = (1\pm \gamma_5)/2$. Since couplings to leptons cannot be directly probed
in a hadron collider environment, we will not concern ourselves with these
in this paper (see \cite{Fox:2011fx} for collider limits on dark matter--electron
couplings). 

In setting bounds we will turn on operators for up and down quarks separately.
The bound for couplings to any linear combination of quark flavors can be derived from these bounds
(see section~\ref{sec:monojet}).
The denomination ``$s$-channel'' or ``$t$-channel'' in equations~\eqref{eq:OV}--\eqref{eq:Og}, refers to
the most straightforward ultraviolet (UV) completions of the respective operators.
For instance, $\mathcal{O}_V$ arises most naturally if dark matter
production in $pp$ collisions proceeds through $s$-channel exchange of
a new heavy gauge boson, and $\mathcal{O}_t$ is most easily obtained if
the production process is $t$-channel exchange of a heavy scalar.
In such a UV completion, $\Lambda$ would be given by $M / \sqrt{g_\chi g_q}$,
where $M$ is the mass of the mediator, $g_\chi$ is its coupling to dark
matter and $g_q$ is its coupling to Standard Model quarks. (The gluon operator
$\mathcal{O}_g$ is somewhat special in this respect since the coupling of a scalar
mediator to two gluons is in itself a dimension-5 operator).  
In supersymmetric theories the dominant interaction of dark matter with quarks is often induced by squark exchange.  For the case of degenerate left and right handed squarks an operator of the form $\mathcal{O}_t$ is predicted (but with $\chi$ being a Majorana fermion).
Here we have assumed that DM is a Dirac fermion, the case of a Majorana fermion~\cite{Rajaraman:2011wf} would not greatly alter our results, except in the case of the vector operator $\mathcal{O}_V$, which vanishes if $\chi$ is a Majorana fermion.  

Ultimately we wish to compare the collider bounds to direct detection bounds, and when matching quark level operators to nucleon level operators the coupling between the SM and DM must be of the form $\mathcal{O}_{\mathrm{SM}}\mathcal{O}_{\chi}$, where $\mathcal{O}_{\mathrm{SM}}$ involves only Standard Model fields and $\mathcal{O}_{\chi}$ involves only dark matter, so that the matrix element $\bra{N}\mathcal{O}_{\mathrm{SM}}\ket{N}$ can be extracted \cite{Fan:2010gt}.  
An operator like $\mathcal{O}_t$, which is not in this form, can be converted into it by a Fierz transformation. This leads to a sum of several operators that can all contribute to the interaction.  Typically, for direct detection, one of these operators will dominate, but at colliders there can be considerable interference. For instance, we can rewrite equation~\eqref{eq:OSt} as
\begin{align}
  \frac{1}{\Lambda^2} (\bar{\chi}P_R q)(\bar{q} P_L \chi) + (L\leftrightarrow R)
    &= \frac{1}{4 \Lambda^2}\left[ (\bar{\chi} \gamma^\mu \chi)(\bar{q} \gamma_\mu q)
     - (\bar{\chi} \gamma^\mu \gamma_5 \chi)(\bar{q} \gamma_\mu \gamma_5 q) \right]
     = \frac{1}{4 \Lambda^2} (\mathcal{O}_V - \mathcal{O}_A) \,.
  \label{eq:Fierz}
\end{align}
If $\chi$ is a Dirac fermion both the $\mathcal{O}_V$ and the $\mathcal{O}_A$ components contribute to $\chi$ production at colliders, but in direct detection experiments, the spin-independent interaction induced by $\mathcal{O}_V$ dominates over the spin-dependent interaction due to $\mathcal{O}_A$. For Majorana dark matter, of course, $\mathcal{O}_V$ would vanish in all cases.

\section{Mono-jets at the LHC}
\label{sec:monojet}

In this section we will derive bounds on dark matter operators with mono-jet searches. In the following subsection we will compare the reach of several mono-jet searches, a low luminosity (36 pb$^{-1}$) CMS search and three ATLAS searches with varying jet $p_T$ cuts using 1 fb$^{-1}$ of data.\footnote{As we were completing this manuscript, CMS has also updated its mono-jet analysis using 1.1 fb$^{-1}$ of data~\cite{CMS-PAS-EXO-11-059}.} For simplicity we will make this comparison only for the vector operator $\mathcal{O}_V$, with dark matter coupling only to up quarks. We will find that the highest jet $p_T$ cuts are most effective in setting bounds on this dark matter interaction. In the next subsection we will proceed to use the analysis based on these highest jet-$p_T$ cuts to set bounds on all effective operators discussed in section~\ref{sec:formalism}.

\subsection{Comparing Various Mono-Jet Analyses}

\begin{figure}
  \begin{center}
    \includegraphics[width=5cm]{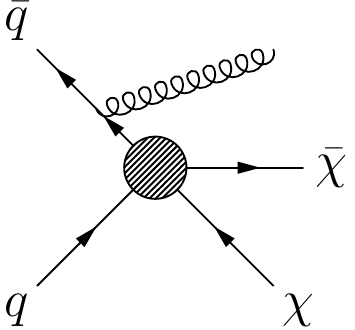}
  \end{center}
  \caption{Dark matter production in association with a single jet in a hadron
    collider.}
  \label{fig:monojet-diagram}
\end{figure}

Dark matter pair production through a diagram like
figure~\ref{fig:monojet-diagram} is one of the leading channels for 
dark matter searches at hadron colliders~\cite{Goodman:2010yf,Bai:2010hh}. The
signal would manifest itself as an excess of jets plus missing energy ($j +
\slashed{E}_T$) events over the Standard Model background, which consists
mainly of $(Z \to \nu\nu) + j$ and $(W \to \ell^{\rm inv} \nu) + j$ final
states.  In the latter case the charged lepton $\ell$ is lost, as
indicated by the superscript ``inv''. Experimental studies of $j +
\slashed{E}_T$ final states have been performed by CDF~\cite{Aaltonen:2008hh},
CMS~\cite{Collaboration:2011nd} and ATLAS~\cite{Collaboration:2011xw,ATLAS-CONF-2011-096}, mostly
in the context of Extra Dimensions.

Our analysis will, for the most part, be based on the ATLAS search~\cite{ATLAS-CONF-2011-096} which looked for mono-jets in 1 fb$^{-1}$ of data, although we will also compare to the earlier CMS analysis~\cite{Collaboration:2011nd}, which used 36 pb$^{-1}$ of integrated luminosity.  The ATLAS search contains three separate analyses based on successively harder $p_T$ cuts, the major selection criteria from each analysis that we apply in our analysis are given below.\footnote{Both ATLAS and CMS impose additional isolation cuts, which we do not mimic in our analysis for simplicity and since they would not have a large impact on our results.}
\begin{description}
\item[\texttt{LowPT}] Selection requires $\met>120$~GeV, one jet with $p_T(j_1)>120$~GeV, $|\eta(j_1)|<2$, and events are vetoed if they contain a second jet with $p_T(j_2)>30$~GeV and $|\eta(j_2)|<4.5$.
\item[\texttt{HighPT}] Selection requires $\met>220$~GeV, one jet with $p_T(j_1)>250$~GeV, $|\eta(j_1)|<2$, and events are vetoed if there is a second jet with $|\eta(j_2)|<4.5$ and with either $p_T(j_2)>60$~GeV or $\Delta\phi(j_2,\met)<0.5$.  Any further jets with $|\eta(j_2)|<4.5$ must have $p_T(j_3)<30$~GeV.
\item[\texttt{veryHighPT}] Selection requires $\met>300$~GeV, one jet with $p_T(j_1)>350$~GeV, $|\eta(j_1)|<2$, and events are vetoed if there is a second jet with $|\eta(j_2)|<4.5$ and with either $p_T(j_2)>60$~GeV or $\Delta\phi(j_2,\met)<0.5$.  Any further jets with $|\eta(j_2)|<4.5$ must have $p_T(j_3)<30$~GeV.
\end{description}
In all cases events are vetoed if they contain any hard leptons, defined for electrons as $|\eta(e)|<2.47$ and $p_T(e)>20$~GeV and for muons as $|\eta(\mu)|<2.4$ and $p_T(\mu)>10$~GeV.

The cuts used by CMS are similar to those of the \texttt{LowPT} ATLAS analysis.  
Mono-jet events are selected by
requiring $\met > 150$~GeV and one jet with $p_T(j_1) > 110$~GeV and
pseudo-rapidity $|\eta(j_1)| < 2.4$. A second jet with $p_T(j_2) > 30$~GeV is allowed if
the azimuthal angle it forms with the leading jet is $\Delta \phi(j_1, j_2) <
2.0$~radians. Events with more than two jets with $p_T > 30$~GeV are vetoed, as
are events containing charged leptons with $p_T>10$ GeV. The number of expected and observed events in the various searches is shown in table~\ref{tab:events}.

\begin{table}
   \centering
   \parbox{14cm}{
   \begin{ruledtabular}
   \begin{tabular}{ccccc}
      & ATLAS \texttt{LowPT} & ATLAS \texttt{HighPT} & ATLAS \texttt{veryHighPT} & CMS \\
      &    1.0 fb$^{-1}$     &      1.0 fb$^{-1}$    &      1.0 fb$^{-1}$        & 36 pb$^{-1}$ \\
      \hline
      Expected & $15100 \pm 700$ & $1010\pm 75$ & $193\pm 25$ & $297 \pm 45$ \\
      Observed & 15740 & 965 & 167 & 275
   \end{tabular}
   \end{ruledtabular}}
   \caption{The expected and observed number of events at ATLAS and CMS, the error is a combination of a) Monte Carlo statistical uncertainties, and b) control sample statistical uncertainties and other systematic uncertainties.  For the case of ATLAS we have combined a) and b) in quadrature.}
   \label{tab:events}
\end{table}

\begin{figure}
  \begin{center}
    \includegraphics[width=0.45\textwidth]{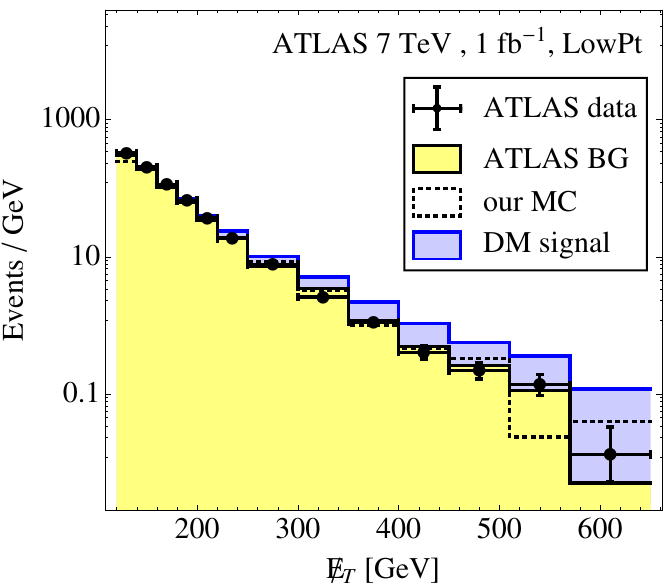}
    \includegraphics[width=0.45\textwidth]{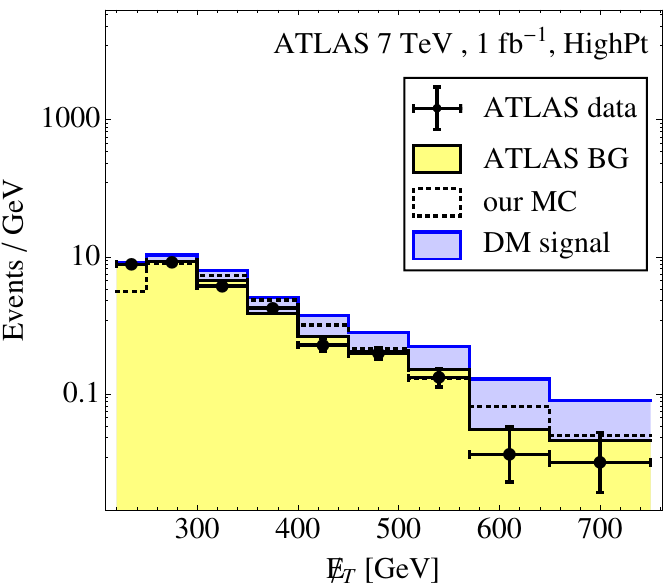} \\[0.3cm]
    \includegraphics[width=0.45\textwidth]{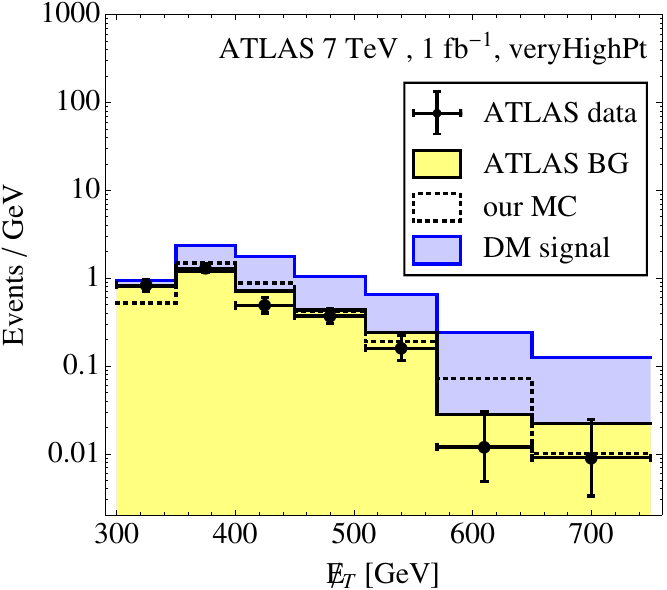} 
    \includegraphics[width=0.45\textwidth]{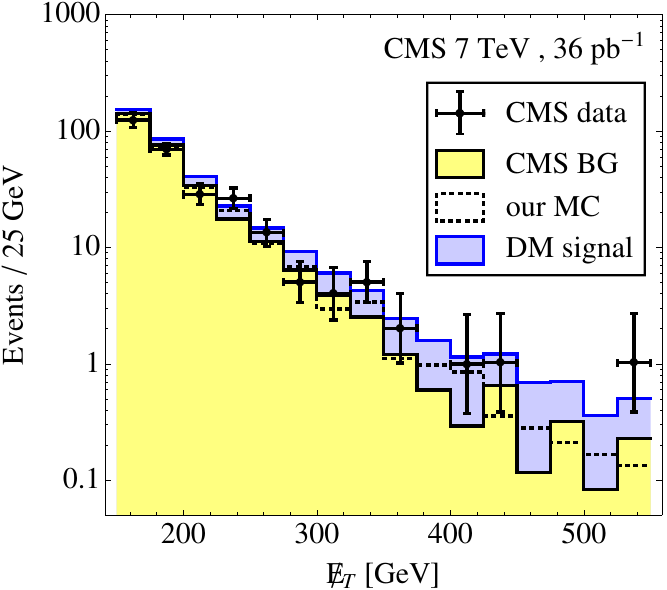}
  \end{center}
  \caption{ Measured missing energy spectra of $j + \met$ for the three ATLAS analyses and the CMS analysis discussed in the text (black data points with error bars) compared to the collaborations' background predictions (yellow shaded histograms) and to our Monte Carlo prediction with (blue histograms) and without (black dotted lines) a dark matter signal.  In all cases the DM signal comes from the vector operator, $\mathcal{O}_V$, and $m_\chi = 10\,\gev$, $\Lambda=400\,\gev$.  Our simulations are rescaled to match the overall normalization of the collaborations' background predictions.}
  \label{fig:CMS-monojet}
\end{figure}

We have simulated the dominant Standard Model backgrounds $(Z \to \nu\nu) + j$ and $(W \to \ell^{\rm inv} \nu) + j$ using MadGraph~\cite{Alwall:2007st,Alwall:2011uj} at the matrix element level, Pythia~6~\cite{Sjostrand:2006za} for parton showering and hadronization, and PGS~\cite{PGS} as a fast detector simulation.  We have checked that results obtained with Delphes~\cite{Ovyn:2009tx} as an alternative detector simulation, would change our results by only a few per cent. In figure~\ref{fig:CMS-monojet}, we compare our simulation of the dominant backgrounds to both the data and the MC predictions of both collaborations\footnote{Note that the MC predictions of the collaborations are for all backgrounds.  For the highest $\met$ bins the background is completely dominated by $W+j$ and $Z+j$, but in the lowest bins there can be $\sim 10\%$ contributions from $t\bar{t}$, QCD and other reducible backgrounds which we did not simulate.}, we also show the spectrum for candidate dark matter models.  In each case we rescale the normalization of the two backgrounds by a correction factor chosen to fit the number of events predicted by the collaborations. After this rescaling  we find excellent agreement in shape between our predictions and theirs.  When predicting the dark matter signal, we rescale the rate by the correction factor found for the invisible $Z$ background, since this background is most similar to the DM signal. The correction factors are approximately 0.9, 1.1 and 1.2 for the three ATLAS analyses (from low to very high respectively), and approximately 0.7 for the CMS analysis. 

As can be seen in figure~\ref{fig:CMS-monojet}, our simulation of Standard
Model backgrounds is in very good agreement with the CMS and ATLAS background
predictions and with the data, so that we can have confidence in our simulations
also for the signal predictions.  Turning to those, we see from
figure~\ref{fig:CMS-monojet} that a dark matter signal mainly changes the slope
of the distribution, leading to the most significant effects at high $\met$~\cite{Beltran:2010ww,Bai:2010hh,Fox:2011fx}.
The main reason for the difference in shape is that dark matter production in
the effective theory framework is a $2 \to 3$ process proceeding through
non-renormalizable operators, whereas the dominant Standard Model backgrounds
have $2 \to 2$ kinematics. 

\begin{figure}
  \begin{center}
    \includegraphics[width=0.5\textwidth]{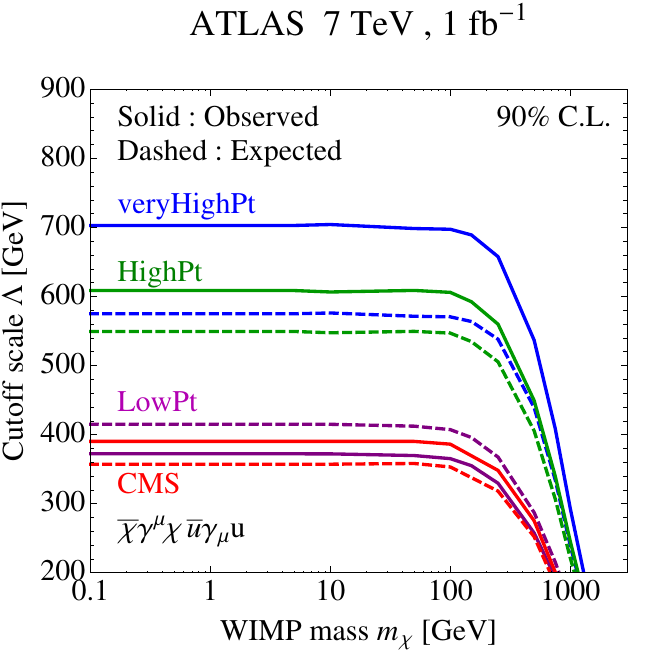}
  \end{center}
   \caption{Limits on the suppression scale $\Lambda$ for the vector operator, $\mathcal{O}_V$, where only the coupling to up quarks is considered, for the three ATLAS analyses and the analysis of CMS.  In all cases the observed (expected) bound is represented by a solid (dashed) line.}
  \label{fig:Lambda-monojet}
\end{figure}

Despite this clear difference in shape between the signal and the background we will nonetheless use only the total event rate to place constraints on dark matter properties since we cannot reliably model systematic uncertainties in the background shape.  However, the existence of three ATLAS analyses with different $p_T$ cuts allows a crude version of a shape analysis to be carried out.  Since the DM signal spectrum is harder than the background spectrum one would expect harder selection cuts to improve the ratio of signal to background, as is reflected in figure~\ref{fig:CMS-monojet}.  To quantify this we compute the expected and observed 90\% exclusion limits on the dark matter--SM coupling, parameterized by the suppression scale $\Lambda$, for a given dark matter mass $m_\chi$ by requiring
\begin{align}
  \chi^2 \equiv \frac{[\Delta_N- N_{\rm DM}(m_\chi, \Lambda)]^2}
                     {N_{\rm DM}(m_\chi, \Lambda) + N_{\rm SM} + \sigma_{\rm SM}^2} = 2.71 \,.
  \label{eq:chi2-totalrate}
\end{align}
Here $\sigma_{\rm SM}$ is the uncertainty in the predicted number of background events, see table~\ref{tab:events}.  In computing the number of expected signal events, $N_{\mathrm{DM}}$, we include the correction factor discussed above to account for the inaccuracy of our detector simulation. We define a quantity 
\be
\Delta_N = \begin{cases} 0 & \text{expected bound} \\
N_{\rm obs} - N_{\rm SM}
& \text{observed bound}~,
\end{cases}
\ee
where $N_{\rm obs(SM)}$ is the number of observed (predicted background) events.  With the exception of the \texttt{LowPT} analysis at ATLAS, all analyses experienced a downward fluctuation in the background and hence give stronger bounds on DM than expected.  Since this lucky accident is unlikely to be repeated in the future we will in the following show both the observed and expected bounds.  The limits on $\Lambda$ for the operator $\mathcal{O}_V$, with coupling to up quarks only, is shown in figure~\ref{fig:Lambda-monojet}.  As expected the strongest bounds come from the analysis with the hardest jet $p_T$ and $\met$ cuts, and in all cases but \texttt{LowPT} the observed bound is stronger than expected due to the downward fluctuations in the data.

It is interesting to note that the CMS and ATLAS \texttt{LowPT} bounds are comparable despite the fact that CMS used 36 pb$^{-1}$ of data whereas ATLAS used 1 fb$^{-1}$.  This is because both analyses are dominated by systematic uncertainties which do not decrease much with luminosity. This clearly illustrates the utility of making cuts that concentrate on the high $p_T$ tail of the mono-jet distribution rather than simply acquiring more luminosity.  The ability to harden cuts and focus on the tails of the distribution increases as the tails get populated with growing luminosity.  Exactly what the best cuts for the DM search are is unclear since there is not much difference between expected bounds from the \texttt{HighPT} and \texttt{veryHighPT} analyses, despite a considerable hardening of cuts.  A dedicated search, with tuned $p_T$ and $\met$ cuts, would presumably lead to even stronger bounds than those coming from ATLAS \texttt{veryHighPT}, we strongly advocate for such a study to be carried out.

The high $p_T$ analyses are most sensitive to the vector operator in the case in which it involves only up quarks. We have also investigated other operators and found that the advantage of the high $p_T$ cuts persists, unless the operator involves only heavier, ``sea", quarks, such as strange or charm. For operators involving these the low $p_T$ analysis does equally well.
The reason is that for sea quarks the parton distribution functions are more rapidly falling, which leads to a softer $p_T$ spectrum more similar to the background spectrum.

Since the expected bounds from the \texttt{HighPT} and \texttt{veryHighPT} analyses are comparable,  we will concentrate from now on on only the \texttt{veryHighPT} ATLAS analysis, and show both the expected and observed bounds from this analysis.  It should be noted that the \texttt{veryHighPT} analysis had the largest fractional downward fluctuation and so the observed bound is considerably stronger than expected, this is unlikely to repeat with more luminosity.  However, exactly how the expected bounds change with luminosity is not straightforward because this depends on  the details of systematic uncertainties at yet higher $p_T$ with higher luminosity.

\begin{figure}[t]
  \begin{center}
    \includegraphics[width=0.45\textwidth]{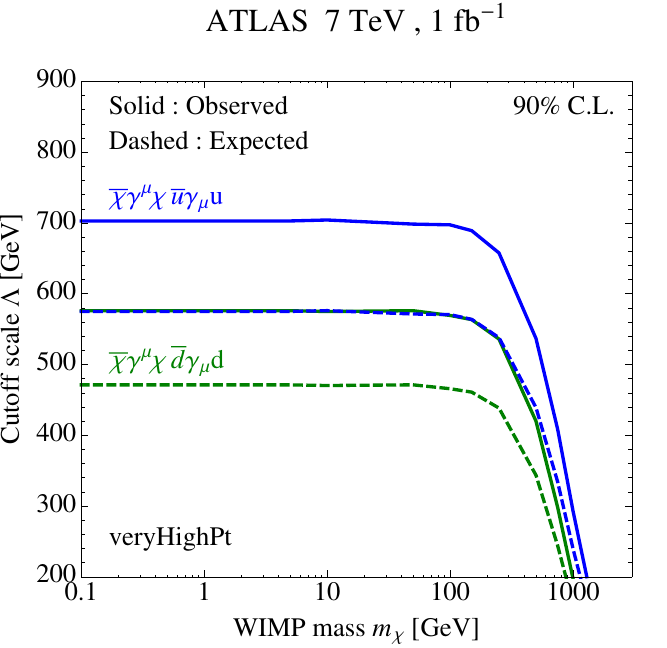}
    \includegraphics[width=0.45\textwidth]{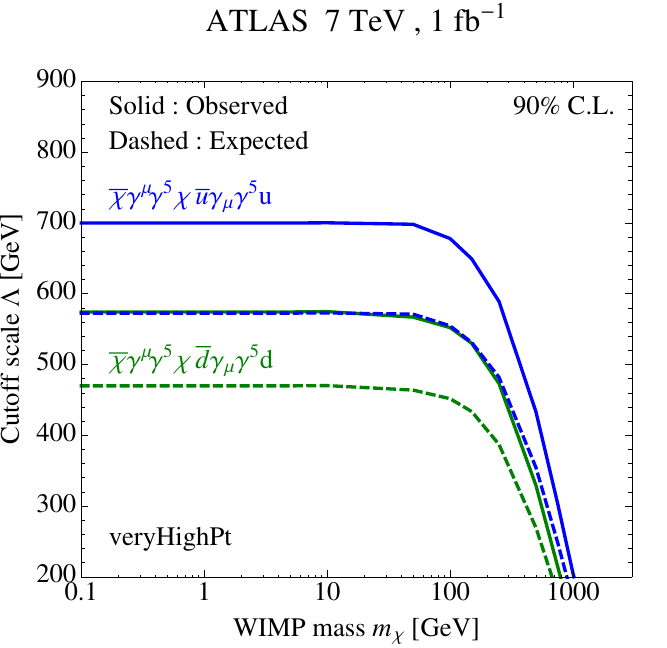}\\[0.5cm]
    \includegraphics[width=0.45\textwidth]{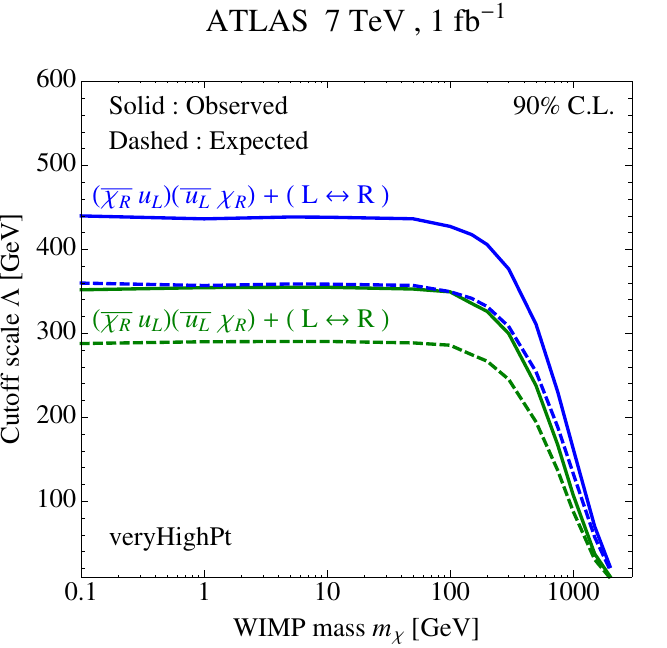}
    \includegraphics[width=0.45\textwidth]{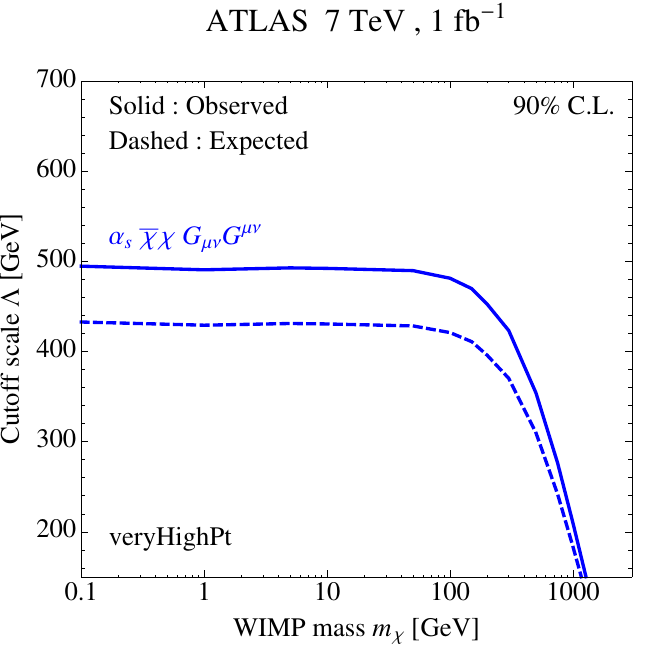}
  \end{center}
   \caption{Limits on the suppression scale $\Lambda$ for various operators, where only the coupling to one quark flavor at a time is considered, for the \texttt{veryHighPT} ATLAS analysis.  In all cases the observed (expected) bounds are shown as solid (dashed) lines.}
  \label{fig:alllambdabounds}
\end{figure}

We can repeat the exercise above for each operator in turn, for both light quark flavors individually.  The results for $\mathcal{O}_V$, $\mathcal{O}_A$, $\mathcal{O}_t$ and $\mathcal{O}_g$ are shown in figure~\ref{fig:alllambdabounds}. As for earlier Tevatron analyses~\cite{Goodman:2010yf,Bai:2010hh}, we note that the collider bounds on the various operators are similar to one another. The collider limits are not strongly affected by the spin structure of the operator, which, as we shall soon see, will give these bounds a relative advantage over direct detection experiments for spin-dependent dark matter scattering typically mediated by axial-vector operators. The bound on the $t$-channel operator $\mathcal{O}_t$ is somewhat weaker than the bound on $\mathcal{O}_V$ and $\mathcal{O}_A$ because of the prefactor $1/4$ and because of partial negative interference between the two terms on the right hand side of equation~\eqref{eq:Fierz}.
The bound on the gluon operator $\mathcal{O}_g$ is very strong, considering that the definition of this operator contains a factor $\alpha_s$, because of the high gluon luminosity at the LHC, despite the operator being of higher dimension than the other operators we consider.

The bounds on the suppression scales of individual operators can be combined for testing models that predict contributions from multiple operators suppressed by the same scale. For instance, consider a model in which dark matter couples to up and down quarks with couplings proportional to $c_u / \Lambda^2$ and $c_d / \Lambda^2$, where $\Lambda$ is a joint suppression scale and $c_u$, $c_d$ are dimensionless coefficients. The constraint on $\Lambda$ can be obtained from the individual constraints on couplings only to up quarks, $\Lambda_u$, and only to down quarks, $\Lambda_d$, from figure~\ref{fig:alllambdabounds} according to the relation
\begin{align}
  \Lambda^4 = c_u^2 \Lambda_u^4 + c_d^2 \Lambda_d^4~.
  \label{eq:flavor-combo}
\end{align}

\subsection{Mono-Jet Bounds Compared to Direct Dark Matter Searches}
\label{sec:monojet-dd}

With these collider bounds in hand we can now place constraints on direct detection rates, in a similar fashion to~\cite{Goodman:2010yf,Bai:2010hh,Goodman:2010ku,Fox:2011fx,Rajaraman:2011wf}. For the coefficients required to translate the quark level matrix elements $\bra{N} \bar{q} \gamma^\mu q \ket{N}$ and $\bra{N} \bar{q} \gamma^\mu \gamma^5 q \ket{N}$ into nucleon level matrix elements, we use the values from \cite{Cheng:1988im,Belanger:2008sj,Barger:2008qd}, as collected in~\cite{Bai:2010hh}. We also need the matrix element for the gluon field strength in the nucleon~\cite{Shifman:1978zn},
\be
\bra{N} \alpha_s G^a_{\mu\nu}G^{a \mu\nu}\ket{N} = -\frac{8\pi}{9} \Big(m_N-\sum_{q=u, d, s} \bra{N}m_q \bar{q} q \ket{N}\Big)~.
\label{eq:MatrixElementGluon}
\ee
For $\bra{N}m_q \bar{q} q \ket{N}$, we follow \cite{Ellis:2008hf} but use an updated \cite{Young:2009ps} value of the pion-nucleon sigma term $\Sigma_{\pi N}=55$ MeV.\footnote{Note however that recent lattice determinations~\cite{Young:2009zb,Toussaint:2009pz,Giedt:2009mr,Takeda:2010cw} of the strange quark content of the nucleon are considerably lower. The effect on our bounds is small.}

When translating collider limits on effective dark matter--Standard Model couplings into constraints on the dark matter--nucleon scattering cross section, we make the simplifying assumption that the couplings are universal in quark flavor.  If flavor-ratios different from unity are desired it is straightforward to translate the collider bounds into direct detection constraints using equation~\eqref{eq:flavor-combo}, with $c_u \neq c_d$. In other words, the LHC limits on the dark matter--nucleon cross section shown in figure~\ref{fig:dd-monojet} would have to be rescaled by a factor $(\Lambda_u^4 + \Lambda_d^4) / (c_u^2 \Lambda_u^4 + c_d^2 \Lambda_d^4)$.

The bounds on the dark matter--nucleon scattering cross sections for the various operators, along with bounds (and some notable excesses) from dedicated direct detection experiments are shown in figure~\ref{fig:dd-monojet}. A few summary comments are in order:
%
\begin{figure}
  \begin{center}
    \includegraphics[width=0.45\textwidth]{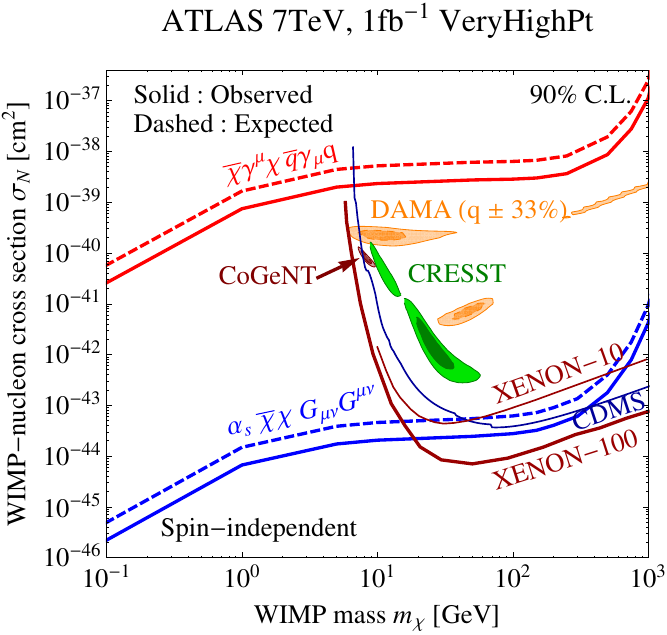} 
    \includegraphics[width=0.45\textwidth]{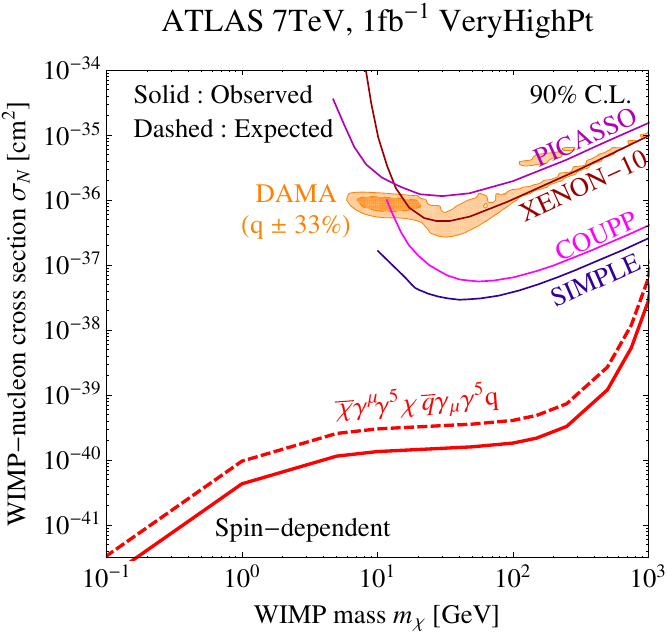}\\
  \end{center}
  \caption{ATLAS limits on (a) spin-independent and (b) spin-dependent dark
    matter--nucleon scattering, compared to limits from the direct detection
    experiments. In particular, we show constraints on spin-independent
    scattering from CDMS~\cite{Ahmed:2009zw}, XENON-10~\cite{Angle:2007uj},
    XENON-100~\cite{Aprile:2011hi}, DAMA~\cite{Bernabei:2008yi},
    CoGeNT~\cite{Aalseth:2011wp,Fox:2011px} and CRESST~\cite{Angloher:2011uu},
    and constraints on spin-dependent scattering from
    DAMA~\cite{Bernabei:2008yi}, PICASSO~\cite{BarnabeHeider:2005pg},
    XENON-10~\cite{Angle:2008we}, COUPP~\cite{Behnke:2010xt} and
    SIMPLE~\cite{Girard:2011xc}. DAMA and CoGeNT allowed regions are based on
    our own fits~\cite{Kopp:2009qt,Fox:2011fx,Fox:2011px} to the experimental
    data. Following \cite{Hooper:2010uy}, we have conservatively assumed large
    systematic uncertainties on the DAMA quenching factors: $q_{\rm Na} = 0.3
    \pm 0.1$ for sodium and $q_{\rm I} = 0.09 \pm 0.03$ for iodine, which leads to
    an enlargement of the DAMA allowed regions. All limits
    are shown at 90\% confidence level, whereas for DAMA and CoGeNT we show
    90\% and $3\sigma$ contours. For CRESST, the contours are $1\sigma$ and $2\sigma$
    as in~\cite{Angloher:2011uu}.}
  \label{fig:dd-monojet}
\end{figure}
%
\begin{itemize}
  \item For spin-independent dark matter couplings, the LHC bounds provide the most powerful constraints for $m_\chi$ below about 5~GeV for the scalar and vector operators and below 10~GeV for the gluon operator.
    
  \item The LHC bound on the vector operator is within 1--2 orders of magnitude from the parameter region suggested by DAMA, CoGeNT and CRESST. The bound on the gluon operator $\mathcal{O}_g$ is several order of magnitude stronger and is competing with CDMS and XENON for dark matter masses up to about 500~GeV.

  \item The LHC provides the strongest bound on spin dependent dark matter--nucleon scattering, by a margin of about two orders of magnitude. The LHC bound becomes less powerful than current direct detection experiments for $m_\chi \gtrsim 1-2$~TeV.
\end{itemize}

\subsection{Limits on Dark Matter Annihilation}

In addition to limits on direct detection cross sections, we have also studied
the constraints that the LHC can set on dark matter annihilation cross sections
relevant to indirect astrophysical searches. The dark matter annihilation rate
is proportional to the quantity $\ev{\sigma v_{\rm rel}}$, where $\sigma$ is
the annihilation cross section, $v_{\rm rel}$ is the relative velocity of the
annihilating particles, and the average $\ev{\cdot}$ is over the dark matter
velocity distribution in the particular astrophysical environment considered.
Working again in the effective field theory framework, we find for dark matter
coupling to quarks through the dimension 6 vector operator, equation~\eqref{eq:OV},
or the axial-vector operator, equation~\eqref{eq:OA}, respectively~\cite{Fox:2011fx},
\begin{align}
  \sigma_V v_{\rm rel} &= \frac{1}{16\pi\Lambda^4} \sum_q \sqrt{1-\frac{m_q^2}{m_\chi^2}} 
    \left(24 (2m_\chi^2+m_q ^2) + \frac{8 m_\chi^4-4m_\chi^2 m_q ^2+5m_q ^4}
                                       {m_\chi^2-m_q ^2}\,v_{\rm rel}^2\right), \\
  \sigma_A v_{\rm rel} &= \frac{1}{16\pi\Lambda^4} \sum_q \sqrt{1-\frac{m_q^2}{m_\chi^2}}
    \left( 24  m_q ^2  +  \frac{8 m_\chi^4-22m_\chi^2 m_q^2+17m_q ^4}
                               {m_\chi^2-m_q ^2}\,v_{\rm rel}^2\right) \,.
\end{align}
Here the sum runs over all kinematically accessible quark flavors, and $m_q$ denotes the quark masses.
We see that, for both types of interaction, the leading term in $\sigma v_{\rm
rel}$ is independent of $v_{\rm rel}$ when there is at least one
annihilation channel with $m_q^2 \gtrsim m_\chi^2 v_{\rm rel}^2$.
Note that for DM
couplings with different Lorentz structures (for instance scalar couplings),
the annihilation cross section can
exhibit a much stronger $v_{\rm rel}$-dependence. For such operators, collider
bounds on $\ev{\sigma v_{\rm rel}}$ can be significantly \emph{stronger} than in
the cases considered here, especially in environments with low $\ev{v_{\rm rel}^2}$
such as galaxies (see, for instance, reference~\cite{Fox:2011fx} for a
more detailed discussion).

\begin{figure}
  \begin{center}
    \includegraphics[width=0.45\textwidth]{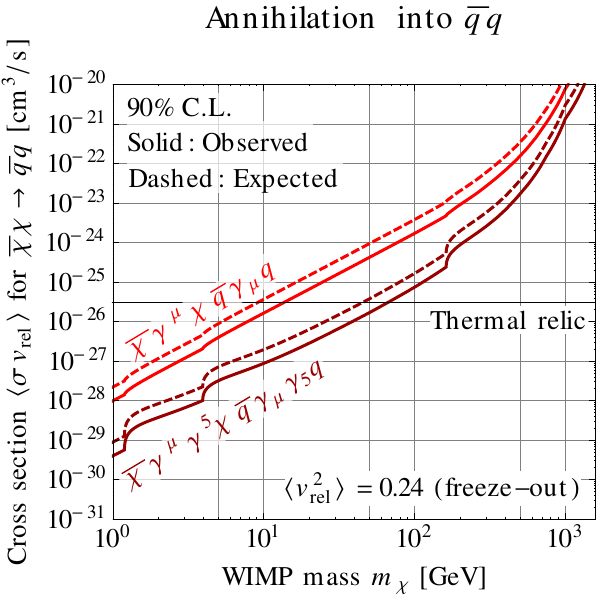}
  \end{center}
  \caption{ATLAS constraints on dark matter annihilation for flavor-universal
    vector or axial vector couplings of dark matter to quarks. (If dark matter
    can annihilate also to leptons, the bounds are weakened by a factor
    $1/{\rm BR}(\bar\chi \chi \to \bar{q} q)$.) We consider an environment
    with $\ev{v_{\rm rel}^2} = 0.24$, corresponding to the epoch at which
    thermal relic dark matter freezes out in the early universe. $\ev{v_{\rm rel}^2}$
    is much smaller in present-day environments such as galaxies,
    which leads to improved collider bounds on the annihilation rate in
    those systems. The value of $\ev{\sigma v_{\rm rel}}$ required for dark
    matter to be a thermal relic is indicated by the horizontal black line.}
  \label{fig:annihilation}
\end{figure}

In figure~\ref{fig:annihilation}, we show ATLAS constraints on $\ev{\sigma v_{\rm rel}}$
as a function of the dark matter mass $m_\chi$ for a scenario in which dark matter
couples equally to all quark flavors and chiralities, but not to leptons.
(If dark matter can annihilate also to leptons, the bounds are weakened by
a factor $1/{\rm BR}(\bar\chi \chi \to \bar{q} q)$.) To compute these limits, we have used
the bounds on $\Lambda_u$ and $\Lambda_d$ from figure~\ref{fig:alllambdabounds},
and have converted them into a limit on the flavor-universal cutoff scale $\Lambda$
using equation~\eqref{eq:flavor-combo}. We have neglected the small contribution
of initial states involving strange and charm quarks to the mono-jet rate
at the LHC.

We see from figure~\ref{fig:annihilation} that, as long as the effective field
theory framework provides a valid description of dark matter production at the LHC
and of its annihilation in the early universe, thermal relic cross sections are ruled
out at 90\% confidence level for $m_\chi \lesssim 15$~GeV in the case of
vector couplings and for $m_\chi \lesssim 70$~GeV in the case of axial vector
couplings. As discussed above, the limits can become somewhat weaker if
additional annihilation channels exist, and stronger in environments with
low $\ev{v_{\rm rel}^2}$.

\section{Light mediators}
\label{sec:lightmediators}

So far, we have worked entirely in the effective field theory framework,
assuming the particles that mediate dark matter--Standard Model interactions to
be much heavier than the typical momentum exchanged in mono-jet events, and the production at colliders to be well approximated by a contact operator.
However, given that the LHC is probing record high scales, particularly for
event samples with hard $p_T$ cuts, it is worthwhile to investigate how the
predictions of the effective theory are modified once a propagating particle is
introduced to mediate the interaction of matter and dark matter.

As discussed in~\cite{Bai:2010hh, Goodman:2010ku, Fox:2011fx, Graesser:2011vj, An:2011ck},
the sensitivity of colliders can
change dramatically in this case, either suppressing or enhancing the signal.
In the case of ``$s$-channel'' operators, resonance effects can enhance the
production cross section once the mass of the $s$-channel mediator is within
the kinematic range and can be produced on-shell. This enhancement is
particularly strong when the mediator has a small decay width $\Gamma$, though
it should be noted that within our assumptions $\Gamma$ is bounded from below
due to the open decay channels to jets and to dark matter.

On the other hand, colliders have a relative disadvantage compared to direct
detection experiments in the light mediator case. The reason is that, from
dimensional analysis, the cross section for the collider production process
$pp \to \bar\chi\chi + X$ scales as,
\begin{align}
  \sigma(pp \to \bar\chi\chi + X) \sim \frac{g_q^2 g_\chi^2}{(q^2 - M^2)^2 + \Gamma^2/4} E^2 \,,
  \label{eq:lm-collider}
\end{align}
where $E$ is of order the partonic center-of-mass energy, $M$ is the mass of
the $s$-channel mediator and $q$ is the four momentum flowing through this
mediator.  At the 7~TeV LHC, $\sqrt{q^2}$ has a broad distribution which is
peaked at a few hundred GeV and falls slowly above.  The mediator's width is
denoted by $\Gamma$, and $g_q$, $g_\chi$ are its couplings to quarks and dark
matter, respectively.  The direct detection cross section, on the other hand,
is approximately
\begin{align}
  \sigma(\chi N \to \chi N) \sim \frac{g_q^2 g_\chi^2}{M^4} \mu_{\chi N}^2 \,,
  \label{eq:lm-dd}
\end{align}
with the reduced mass $\mu_{\chi N}$ of the dark matter and the target nucleus.

When $M^2 \ll q^2$, the limit that the collider sets on $g_{\chi}^2 g_q^2$
becomes independent of $M$, whereas the limit on $g_{\chi}^2 g_q^2$ from direct
detection experiments continues to become stronger for smaller $M$. In other
words, the collider limit on $\sigma(\chi N \to \chi N)$ becomes weaker as $M$
becomes smaller. On the other hand, when $m_\chi < M/2$ and the
condition $\sqrt{q^2} \simeq M$ can be
fulfilled, collider production of $\bar\chi\chi + X$ experiences resonant
enhancement. Improved constraints on $\Lambda$ can be expected in that
regime.  

\begin{figure}
  \begin{center}
    \includegraphics[width=0.45\textwidth]{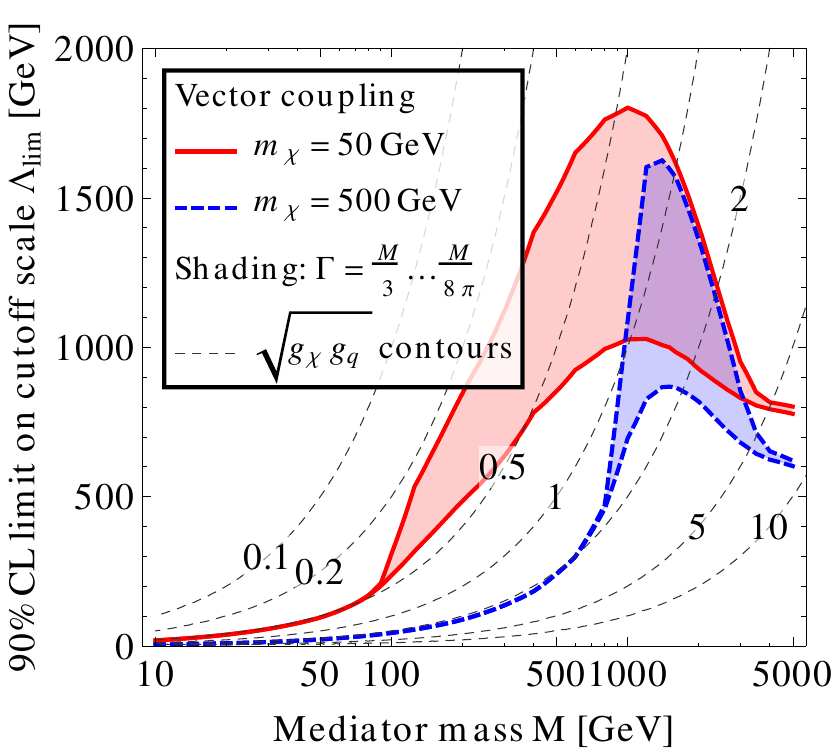}
  \end{center}
  \caption{ATLAS limit on $\Lambda \equiv M / \sqrt{g_{\chi} g_q}$
    as a function of the mass $M$ of the particle mediating dark
    matter--quark interactions. We have assumed $s$-channel vector-type interactions,
    and we have considered the values $m_\chi = 50$~GeV (red) and
    $m_\chi = 500$~GeV (blue) for the dark matter mass. We have varied
    the width $\Gamma$ of the mediator between the values $M/3$ (lower boundary
    of colored bands) and $M/8\pi$ (upper boundary of colored bands).
    Dashed dark gray lines show contours of constant $\sqrt{g_\chi g_q}$.}
  \label{fig:light-mediator}
\end{figure}

In figure~\ref{fig:light-mediator}, we investigate the dependence of the ATLAS
bounds on the mediator mass $M$ more quantitatively including both on-shell and
off-shell production. Even though dark matter--quark interactions can now no
longer be described by effective field theory in a collider environment, we
still use $\Lambda \equiv M / \sqrt{g_{\chi} g_q}$ as a measure for the
strength of the collider constraint, since $\Lambda$ is the quantity that
determines the direct detection cross section. As before, we have used the cuts
from the ATLAS {\tt veryHighPt} analysis (see section~\ref{sec:monojet}).  We
have assumed vector interactions with equal couplings of the intermediate
vector boson to all quark flavors.

At very large $M$ ($\gtrsim 5$~TeV), the limits on $\Lambda$ in
figure~\ref{fig:light-mediator} asymptote to those obtained in the effective
theory framework. For $2m_\chi \ll M \lesssim 5$~TeV, resonant enhancement
leads to a significant improvement in the limit since the mediator can now be
produced on-shell, so that the primary parton--parton collision now leads to a
two-body rather than three-body final state. As expected from
equation~\eqref{eq:lm-collider}, the strongest enhancement occurs when the
mediator is narrow.  In figure~\ref{fig:light-mediator}, we show the effects of
resonance enhancement.  We consider mediators of fixed width, ranging from
$\Gamma = M / 8\pi$ to $\Gamma = M /3$, the associated enhancments are
illustrated by the colored bands, with the upper edge corresponding to the
narrow case and the lower edge to a wide mediator.\footnote{$\Gamma = M / 8\pi$
corresponds to a mediator that can
annihilate into only one quark flavor and helicity and has couplings $g_\chi
g_q = 1$.  Since in figure~\ref{fig:light-mediator}, we have assumed couplings
to all quark helicities and flavors (collider production is dominated by
coupling to up-quarks though), and since $g_\chi g_q > 1$ in parts of the plot
(see dashed contours), $\Gamma = M/8\pi$ can be regarded as an approximate lower limit on
the mediator width.} The shape of the peaks in figure~\ref{fig:light-mediator}
is determined by the interplay of parton distribution functions, which suppress
the direct production of a heavy mediator, and the explicit proportionality of
$\Lambda$ to $M$ according to its definition. Below $M \simeq 2 m_\chi$, the
mediator can no longer decay to $\bar\chi \chi$, but only to $\bar{q} q$, so in
this mass range, it can only contribute to the mono-jet sample if it is
produced off-shell. In that regime, the limit on $\Lambda$ is rather weak (even
though the limit on $g_\chi^2 g_q^2$ is independent of $M$ there as discussed
above), and the dependence on $\Gamma$ disappears.

In light of this result it is important to revisit our limits from
section~\ref{sec:monojet} and check that they are consistent with the effective
theory in which they were derived. In other words, we have to verify that
models which saturates our limits can still be described in effective field
theory.  Inspecting the dashed contours of constant mean coupling $\sqrt{g_q
g_\chi}$ in figure~\ref{fig:light-mediator}, we see that for mediator masses
above $\sim 5$~TeV, where the limits derived in the full renormalizable theory
asymptote to those derived in the effective theory, our limits would
correspond to $\sqrt{g_qg_\chi} \sim 5$--$10$, depending on $m_\chi$. This is
still below the $\sqrt{g_q g_\chi} = 4\pi$, which for
small $m_\chi$ would be reached at $M \sim 10$~TeV. We thus see that there is
considerable parameter space available in the renormalizable model in which
effective theory provides a good low-energy approximation. Moreover, we have
seen that even for lighter mediators, $M \sim \text{few} \times 100$~GeV, the
limits derived from the effective theory are valid, though overly conservative.  However, for very light mediators, $M \lesssim 100$~GeV, the collider bounds on direct detection cross sections are considerably weakened.

Even though we have only quantitatively demonstrated the above conclusions for
dark matter with vector couplings here, the results of references~\cite{Bai:2010hh,Fox:2011fx}
show that they can be generalized to other types of effective operators, in particular
axial vector $\mathcal{O}_A$ and scalar $t$-channel $\mathcal{O}_t$.
For the gluon operator $\mathcal{O}_g$, we remark that its most natural
UV-completion is through a diagram in which the two gluons as well as a new
scalar $s$-channel mediator couple to a triangular heavy quark loop.
Due to the additional loop factor which need not be present in UV completions
of $\mathcal{O}_V$ and $\mathcal{O}_A$, the masses of the new heavy scalar
and the new heavy quark propagating in the loop cannot be larger than $\sim 1$~TeV
for a theory that saturates our limit $\Lambda \sim 500$~GeV (see
figure~\ref{fig:alllambdabounds}). Therefore, as one can see from
figure~\ref{fig:light-mediator}, effective field theory is not strictly
applicable in such a model, but the limit it gives is on the conservative side.

Let us finally comment on the case of scalar dark matter--quark couplings
of the form
\begin{align}
  \mathcal{O}_S \equiv \frac{(\overline{\chi_L} \chi_R)(\overline{q_L} q_R)}{\Lambda^2}
                    + (L\leftrightarrow R)~,
  \label{eq:OS}
\end{align}
which we have not considered so far in this paper. As any UV completion of that
operator has to preserve $SU(2)$ invariance, it is necessary that one
of the chirality eigenstates $\chi$ is an $SU(2)$ doublet or that the UV completion
of $\mathcal{O}_S$ involves coupling to the Higgs field $H$. The first possibility
is strongly constrained because dark matter charged under $SU(2)$ would have
been detected already in direct detection experiments, unless it is very light,
$m_\chi \lesssim \text{few GeV}$. The second possibility, a Higgs insertion,
implies that $\mathcal{O}_S$ should be rewritten as
\begin{align}
  \mathcal{O}_S' \equiv
    \frac{y_q (\overline{\chi_L} \chi_R)(\overline{q_L} \ev{H} q_R)}{\Lambda'^3}
                          + (L\leftrightarrow R)~,
  \label{eq:OSprime}
\end{align}
where $y_q$ is the Standard Model Yukawa coupling of $q$ and $\Lambda'$ is
the cutoff scale of the effective theory (the scale $\Lambda$ from
equation~\eqref{eq:OS} has no physical meaning in the case of a Higgs insertion).
The simplest possibility to realize $\mathcal{O}_S'$ at the renormalizable level is through
mixing of the Higgs with a new scalar singlet, which in turn couples to dark matter.
In this case, both dark matter production at the LHC and dark matter--nucleus
scattering in direct detection experiments are dominated by sea quark
contributions, due to Yukawa suppression. We have checked that, in this case, the
limit the LHC could set on $\Lambda'$ is below 100~GeV
and thus clearly outside the regime of validity of effective field theory.
We will therefore not consider operators of the form $\mathcal{O}_S$ or
$\mathcal{O}_S'$ any further in this paper.

To conclude this section, let us emphasize that here we have only considered
one possible UV completion of the effective operators introduced in section~\ref{sec:formalism}.
While this helps outline some of the main effects of finite mediator masses,
the exact details of these effects will be model-dependent.

\section{Mono-photons at the LHC}
\label{sec:monophoton}

While mono-jets are certainly an excellent search channel for dark matter, it
is important to investigate other complementary channels with different
systematic uncertainties. An interesting final state to consider is the
mono-photon channel, which we will study in this section. A search in an
independent channel can help determine if any excess seen in $j +
\slashed{E}_T$ is due to new physics or due to mismodelling of backgrounds.
Also, there are many types of new physics besides dark matter
that can lead to mono-jet signatures, for instance large extra
dimensions~\cite{ArkaniHamed:1998rs} and unparticles~\cite{Georgi:2007ek}, so
that searches in additional channels will be necessary to narrow down the
origin of any observed signal. In addition, information from several channels
may shed light on the nature of the DM--SM coupling. For example, the relative
size of an excess in mono-photons compared to one in mono-jets is sensitive to
whether the operator dominating the signal involves up or down quarks, due to
their different electric charges. A gluon operator like $\mathcal{O}_g$ from
equation~\eqref{eq:Og} is not expected to produce a significant mono-photon signal at all.

Studies of the mono-photon final state have been carried out by
CDF~\cite{Aaltonen:2008hh} and D\O~\cite{Abazov:2008kp}, but here we follow the
recent CMS analysis, based on 1.14 fb$^{-1}$ of
luminosity~\cite{CMS-PAS-EXO-11-058}.  Single photons can be produced in
association with a dark matter pair through diagrams similar to
figure~\ref{fig:monojet-diagram}, but with the outgoing gluon replaced by a
photon.  Thus, the cross section for mono-photon production is suppressed
compared to mono-jet production by the ratio of the strong and electromagnetic
fine structure constants as well as a color factor. On the other hand, the
background is similarly smaller. Systematic uncertainties on the background
prediction are similar, of order 10--15\%, for the ATLAS {\tt veryHighPT}
mono-jet search and for the CMS mono-photon search~\cite{CMS-PAS-EXO-11-058}.
The acceptance for mono-photons is somewhat lower than that for mono-jets because of the requirement that they fall in the barrel part of the electromagnetic calorimeter. 

In our simulations, we follow~\cite{CMS-PAS-EXO-11-058} and require the photon
to a have transverse momentum $p_T(\gamma) > 95$~GeV and pseudo-rapidity
$|\eta| < 1.44$.  The missing energy in the event must satisfy $\met> 80$~GeV
and the event is vetoed if there is a jet with $p_T(j)>20\ \gev$ within
$|\eta(j)|<3$ or a lepton with $p_T(\ell)>10\ \gev$ and $\Delta
R(\ell,\gamma)>0.04$. CMS applies several additional photon identification and
isolation criteria which we do not attempt to mock up.  Instead, we use PGS as
a detector simulation and apply a correction factor of 0.71 to account for these
isolation requirements. The correction factor is obtained by comparing our
prediction for the dominant irreducible background $(Z \to \nu\nu) + \gamma$
to the collaboration's.  We have also checked that the shape we predict for
$(Z \to \nu\nu) + \gamma$ is in excellent agreement with their
prediction, which provides a useful verification of our simulations.
Apart from $(Z \to
\nu\nu) + \gamma$, the backgrounds in the $\gamma + \slashed{E}_T$ channel are $(Z \to \nu\nu) + j$, with the jet mistaken for a photon, $W \to e\nu$, with the electron mistaken for a photon, bremsstrahlung from cosmic ray or beam halo muons and $(W \to \ell^{\rm
inv}\nu) + \gamma$, with an unidentified charged lepton $\ell$.  The expected
number of events in the mono-photon sample, according to CMS, is $67.3\pm8.4$ (with the
uncertainty dominated by statistics) and the number of observed events was 80.

\begin{figure}[t]
  \begin{center}
    \includegraphics[width=0.45\textwidth]{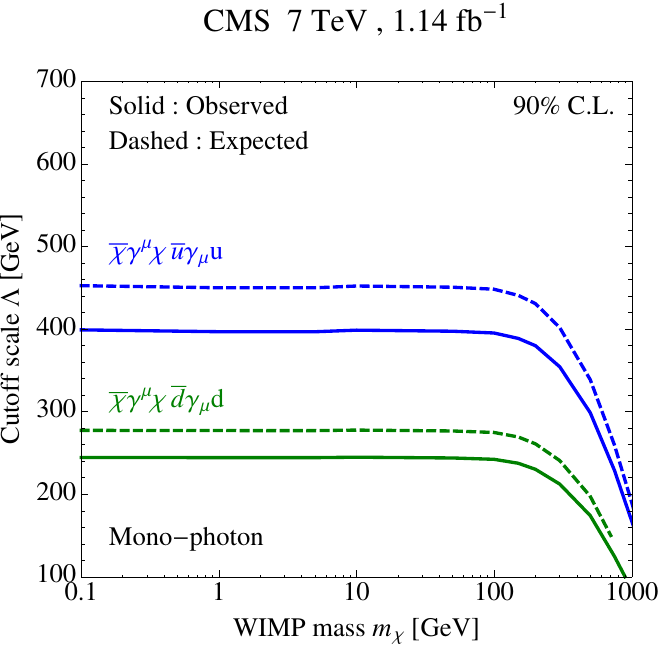}
  \end{center}
   \caption{Limits from the CMS mono-photon analysis on the suppression scale $\Lambda$ for the vector operator, $\mathcal{O}_V$, where only the coupling to one quark flavor at a time is considered.  The expected bound is shown with a dashed line and the observed one with a solid line.}
  \label{fig:alllambdaboundsphoton}
\end{figure}

\begin{figure}
  \begin{center}
    \includegraphics[width=0.45\textwidth]{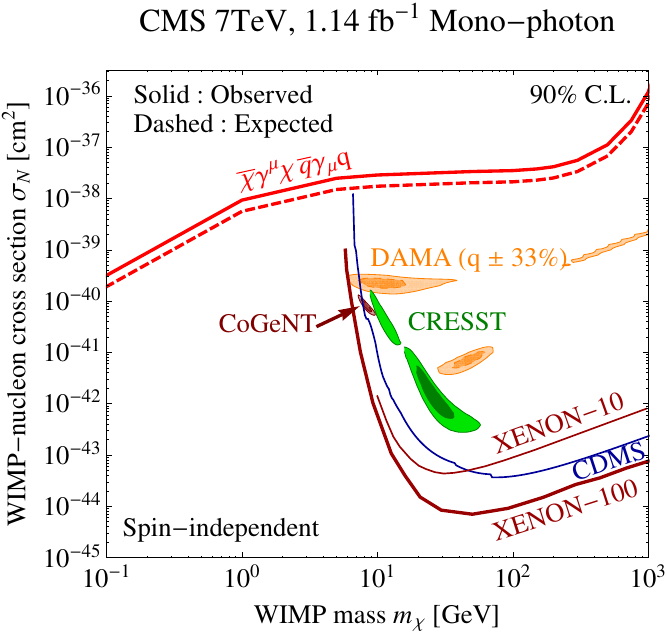} 
  \end{center}
  \caption{Limits from the CMS mono-photon analysis on spin-independent dark
    matter--nucleon scattering, compared to limits from direct detection
    experiments. In particular, we show constraints on spin-independent
    scattering from CDMS~\cite{Ahmed:2009zw}, XENON-10~\cite{Angle:2007uj},
    XENON-100~\cite{Aprile:2011hi}, DAMA~\cite{Bernabei:2008yi},
    CoGeNT~\cite{Aalseth:2011wp,Fox:2011px} and CRESST~\cite{Angloher:2011uu}.
    DAMA and CoGeNT allowed regions are based on
    our own fits~\cite{Kopp:2009qt,Fox:2011fx,Fox:2011px} to the experimental
    data. Following \cite{Hooper:2010uy}, we have conservatively assumed large
    systematic uncertainties on the DAMA quenching factors: $q_{\rm Na} = 0.3
    \pm 0.1$ for sodium and $q_{\rm I} = 0.09 \pm 0.03$ for iodine, which leads to
    an enlargement of the DAMA allowed region. All limits
    are shown at 90\% confidence level, whereas for DAMA and CoGeNT we show
    90\% and $3\sigma$ contours. For CRESST, the contours are $1\sigma$ and $2\sigma$
    as in~\cite{Angloher:2011uu}.}
  \label{fig:dd-monophoton}
\end{figure}

To set limits on dark matter, we add our signal prediction to the number of predicted background events from~\cite{CMS-PAS-EXO-11-058} and compare the result to the CMS data following the same statistical procedure as in section~\ref{sec:monojet}. The resulting limits on the cutoff scale $\Lambda$ for vector operators involving up and down quarks are shown in figure~\ref{fig:alllambdaboundsphoton}. The current mono-photon bounds still trail behind mono-jet limits. However, the mono-photon limits may improve more rapidly than those from mono-jets because the former are still statistics dominated as opposed to the latter which are already dominated by systematic uncertainties. Furthermore, as we saw in the previous section, applying harder $p_T$ cuts may yield stronger bounds. The resulting limit on the direct detection cross section is shown in figure~\ref{fig:dd-monophoton}.

\section{Dark Matter Coupling through Higgs Exchange}
\label{sec:invhiggs}

One of the most motivated scenarios for dark matter is the case where dark matter interacts through the exchange of a Higgs boson~\cite{Burgess:2000yq}. In this section we will consider this possibility. For concreteness we will assume a specific model, the Standard Model plus a dark matter particle that couples via the Higgs ``portal''. We will place limits on the direct detection signal in this model at the LHC in two ways. First, using potential future limits on the invisible branching fraction of the Higgs, we place an future upper bound on the direct detection signal. Then we will use current Higgs limits and assume that the decay of a Higgs to dark matter is responsible for the Higgs non-discovery. This will lead to interesting \emph{lower bounds} on dark matter--nucleon scattering rates.

\subsection{The Invisible Higgs Analysis as a Dark Matter Search}

One way to search for dark matter coupled through the Higgs is to follow the strategy of the previous sections. Namely, integrating out the Higgs induces a scalar operator $\sim (\bar \chi \chi)(\bar q q)$, which is suppressed by the Yukawa couplings, and an operator like $\mathcal{O}_g$ that couples dark matter to gluons after a top quark loop is integrated out. One can then look for a mono-jet (or mono-photon) signal to constrain the magnitude of the operator. However, for the light generations the Yukawa suppression will make the bound on such operators weaker (at least as compared with operators that are not Yukawa suppressed). Furthermore, a light Standard Model Higgs is a ``light mediator'' in the sense that its propagator may easily be dominated by the momentum transfer $q^2$, rather than the mass $m_h^2$, which will lead to another disadvantage of mono-jet searches (see section~\ref{sec:lightmediators}).

Here we will pursue a different strategy which will give stronger bounds within a model in which DM couples via a Higgs boson $h$, particularly when dark matter is so light that the decay $h \to \bar\chi \chi$ is kinematically allowed. Production of a Higgs at the LHC may proceed through the Higgs' gauge, rather than its Yukawa, couplings. In particular, one can produce a Higgs in association with a $Z$ or a $W$ or through vector boson fusion (VBF). If $m_\chi < m_h/2$, the Higgs may have a sizeable branching fraction into missing energy, leading to invisible Higgs signals such as $Z + \met$ (from associated production) or forward jets plus $\met$ (from VBF). For a given Higgs mass, the limits on the invisible branching fraction of the Higgs may be translated into limits on the coupling of the Higgs to dark matter and thus into a limit on the direct detection cross section mediated by a Higgs.\footnote{Some related work on the application of the invisible Higgs search to the dark matter interaction has been discussed in \cite{Kanemura:2010sh, Kanemura:2011nm, He:2011fk}. The bounds on the invisible Higgs branching fraction from XENON-100 in the scalar dark matter case are discussed in \cite{Mambrini:2011ik}}

For concreteness we consider a toy model in which a new neutral and stable dark matter fermion, $\chi$, is added to the Standard Model, coupling to the Higgs.\footnote{One could easily apply our methods also to the case of a minimal model of scalar dark matter~\cite{Burgess:2000yq}, giving similar results, or to models with extended Higgs sectors in which Higgs production can be modified.} For example, this coupling may be written as $\bar \chi  \chi H^\dagger H$, which below electroweak symmetry breaking leads to a coupling of the form $y_\chi h \bar \chi \chi$. In these expressions $H$ denotes the SM Higgs doublet, $h$ stands for the physical Higgs boson, and $y_\chi$ is a dimensionless coupling constant. The branching fraction of the Higgs into dark matter pairs is 
\begin{equation}\label{Brinv}
  \mathrm{BR}(h \to \bar\chi\chi) = 
    \frac{\Gamma(h \to \bar\chi\chi)}{\Gamma(h \to \bar\chi\chi)+\Gamma(\rm{SM})},
    \qquad\Gamma(h \to \bar\chi\chi) = \frac{y_{\chi}^2}{8\pi} \, 
                        m_h\left[1 - \left(\frac{2m_{\chi}}{m_h}\right)^{2}\right]^{3/2}\,,
\end{equation}
where $\Gamma(\rm{SM})$ is the total width of the Higgs in the Standard Model, which depends on the Higgs mass, and $\Gamma(h \to \bar\chi\chi)$ is the partial width for decays into dark matter. The invisible Higgs search from colliders sets an upper bound on $\mathrm{BR}(h \to \bar\chi\chi)$, which in our model constrains the size of $y_\chi$. 
We can then translate this bound into a bound on the direct detection cross section using the couplings of the Higgs to the nucleus at low energies. This can proceed in two ways---the Higgs can couple to the strange quark in the nucleus or it can couple to gluons via a heavy quark loop. These couplings are suppressed either by the Yukawa coupling of the strange quark or by a loop factor, which will give the collider limits a relative advantage since those involve order 1 couplings. 
We use the matrix element for the gluon coupling given in equation~\eqref{eq:MatrixElementGluon} and for the strange quark coupling as discussed in section~\ref{sec:monojet-dd}. The resulting direct detection cross section is 
\begin{equation}\label{CrossInv}
\sigma_{N}=5\times 10^{-6}\,\frac{\mu^2\,y_{\chi}^2}{\pi\,m^4_h}\,,
\end{equation}
which sets the direct detection bounds once we extract the allowed size of $y_{\chi}$ from the invisible Higgs search.  

There are many works discussing the future bounds on invisible Higgs decays~\cite{Gagnon, Davoudiasl:2004aj, Meisel, Gagnon2}. Here we will not conduct a study of our own, but rather take the bounds projected in an ATLAS analysis~\cite{Gagnon2} where the production modes $ZH$ and VBF are considered.
The dominant SM backgrounds for these processes are $ZZ\to\ell\ell\nu\nu$ for the $ZH$ production mode and jets from QCD, $W^{\pm}$ or $Z$ for the VBF case. The authors of~\cite{Gagnon2} have simulated both signal and background with the full ATLAS detector simulation. The systematic uncertainties from Monte Carlo, experimental systematic uncertainty, and the theoretical knowledge of the production cross-sections are taken into account. 

\begin{figure}
  \begin{center}
    \includegraphics[height=0.45\textwidth]{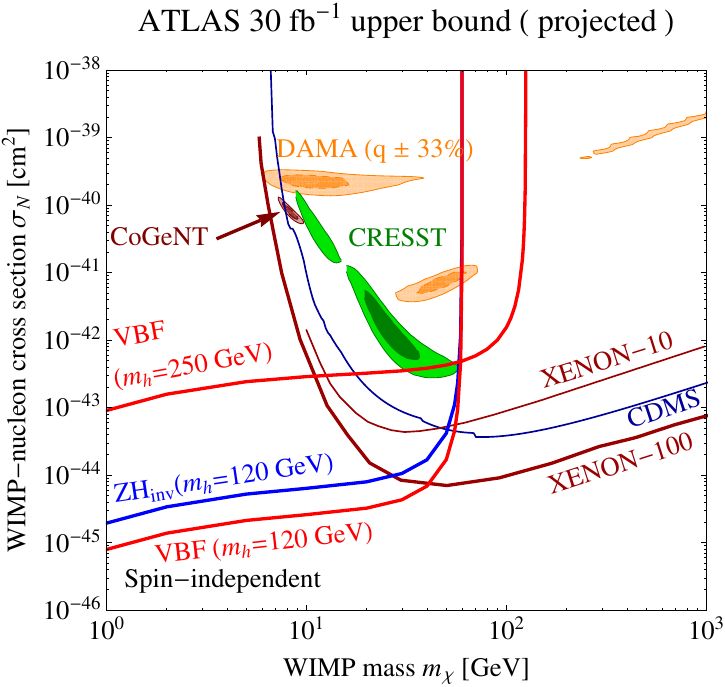}
    \includegraphics[height=0.45\textwidth]{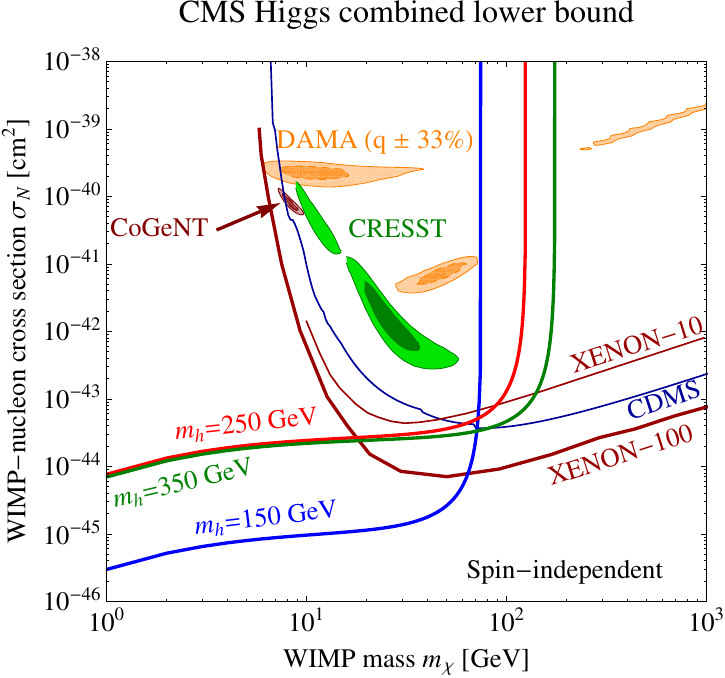}
  \end{center}
  \caption{
    Left: Projected 95\% C.L.\ upper bounds on dark matter--nucleon scattering mediated by a Higgs boson
    from future ATLAS searches for invisible Higgs decays. Limits are shown for the
    $Z+H$ and vector boson fusion (VBF) production modes, and for Higgs masses of
    120~GeV and 250~GeV~\cite{Gagnon2}. Right: \emph{Lower} 95\% C.L.\ bounds on dark
    matter--nucleon scattering mediated by a Higgs boson, derived from the CMS exclusion
    of a Standard Model Higgs boson in certain mass ranges~\cite{CMS-PAS-HIG-11-022},
    assuming that the Higgs was missed at the LHC due to its large invisible width.
    The direct detection limits we show for comparison are the same as in
    figures~\ref{fig:dd-monojet} and \ref{fig:dd-monophoton}.}
  \label{fig:invisibleHiggs}
\end{figure}

Assuming $30\ \rm{fb}^{-1}$ of data with 14~TeV center of mass energy, the projected 95\% C.L.\ upper bounds on the invisible branching ratio are~\cite{Gagnon2}
\begin{center}
  \parbox{6cm}{
  \begin{ruledtabular}
    \begin{tabular}{ccc}
      channel &  $ZH_\mathrm{inv}$  & VBF \\
      \hline
      $m_h=120$ GeV & 0.75 & 0.55  \\
      $m_h=250$ GeV & --   & 0.85  \\
    \end{tabular}
  \end{ruledtabular}}
\end{center}
Using these bounds and equation~\eqref{Brinv}, we can set upper limits on the direct detection cross section. These limits are shown in the left panel of figure~\ref{fig:invisibleHiggs} for various Higgs masses and production channels. These dark matter--nucleon scattering cross section bounds are more stringent than the mono-jet and mono-photon bounds of the previous sections due to the smallness of the Higgs--nucleon coupling. The bounds deteriorate when the dark matter mass approaches the kinematic limit for invisible Higgs decay at $m_\chi = m_h/2$. Comparing the results for different Higgs masses, the bound for a 250~GeV Higgs is weaker than the one for $m_h = 120$~GeV because at 120~GeV, the SM Higgs width $\Gamma(\rm{SM})$ is small, allowing the invisible channel to compete even for moderate couplings. At 250~GeV, the SM decay rate is dominated by decays to $W$ and $Z$ bosons, and in order for the Higgs to have a sizeable invisible branching fraction, the coupling to dark matter must be quite large. This effect over-compensates the $1/m_h^4$ suppression in the direct detection cross section which pushes the limits in the opposite direction. 

\subsection{A Lower Bound on Dark Matter--Nucleon Scattering from Current Higgs Limits}

In the previous subsection we discussed the future reach of the LHC in discovering 
dark matter ``directly'' through invisible Higgs decay. But if dark matter indeed couples to the Standard Model through Higgs exchange, there is always an interesting connection between the Higgs search and the search for dark matter. This is true both for bounds on the Higgs, as well as for a potential Higgs discovery. 

For example, the recent LHC exclusions~\cite{CMS-PAS-HIG-11-022,ATLAS-CONF-2011-135} of a SM Higgs between $\sim 140~\gev$ and $\sim~400~\gev$ have an amusing interpretation as a possible \emph{lower} bound on the dark matter scattering rate expected at direct detection experiments.  In particular, if the Higgs has a sizeable branching fraction into dark matter, this leads to a suppression of the decay channels used in the SM Higgs searches. Thus, a Higgs mass that is inconsistent with data for SM branching ratios may be allowed if the invisible width is large enough to sufficiently suppress the SM search modes, dominantly $h \to W^+ W^-$ or $h \to ZZ$ in the Higgs mass range of interest.

For concreteness we consider the combined Higgs bound from CMS~\cite{CMS-PAS-HIG-11-022}, but results would be similar for the ATLAS bound~\cite{ATLAS-CONF-2011-135}.  Over the mass range $140\ \gev\lesssim m_h \lesssim 400\ \gev$ the bound on $\xi = (\sigma \times {\rm BR})/(\sigma\times {\rm BR})|_{SM}$ varies from $\sim 0.3$--1, here BR is the branching ratio into the relevant search mode, in this mass range either $h \to W^+ W^-$ or $h \to ZZ$.  Using \eqref{Brinv} this can be translated into a lower bound on $y_\chi$ under the assumption that the Higgs is produced with SM cross section, and one decay mode dominates the bound, but Higgs decays into SM channels are suppressed by a large invisible width (and by nothing else),
\be
y_\chi^2 \ge 8\pi \frac{\Gamma_{\rm tot}^{\rm SM}}{m_h} \frac{1-\xi}{\xi} \left(1-\frac{4m_\chi^2}{m_h^2}\right)^{-3/2}~.
\ee
This lower bound on the Higgs--DM coupling allows us to place a \emph{lower} bound on dark matter--nucleon scattering due to Higgs exchange, for a light Higgs that was missed at the LHC due to a large invisible width.  This is shown for several candidate Higgs masses in figure~\ref{fig:invisibleHiggs}. It is interesting to note that in some cases shown (e.g. Higgs masses of 250 and 350 GeV) this lower bound is already in conflict with direct detection limit for a wide range of dark matter masses. This implies that such minimal models of Higgs-coupled dark matter are already being probed by the combination of the LHC and direct detection. To evade these limits non-minimal models must be considered, either modifying Higgs production rates or modifying Higgs decay beyond the dark matter channel.  It will be interesting to follow upcoming limits on the Standard Model Higgs which will cause these lower bounds to rise and possibly come in to conflict either with dark matter searches, or with invisible Higgs searches.

Finally, we note that even if a Standard Model-like Higgs is discovered at the LHC, interesting bounds on direct detection may be extracted. The strength of these bounds as well as their nature, upper or lower, depend on the details of the discovery. For example, assume a Standard Model-like Higgs is discovered at 120~GeV, say in the $\gamma\gamma$ decay mode. If the Higgs production rate times branching fraction agrees with the Standard Model prediction, very little room will be left for decay of the Higgs into light dark mater. Because the decay channel that is competing with dark matter for this Higgs mass, $h \to b \bar b$, has very small branching ratio, this will set a strong upper bound on the coupling of the Higgs to $\chi$ (of order the bottom Yukawa coupling), and thus on direct or indirect detection. 

One the other hand, if the Higgs is discovered with a rate that is below the Standard Model prediction, one can postulate that the decay into dark matter is responsible for the reduced rate. Within this assumption, both an upper and a lower limit on dark matter couplings may be derived. In this case the invisible Higgs search can confirm these assumption and provide a potential dark matter discovery.

\section{Conclusions}
\label{sec:conclusions}

Missing energy signatures have long been known to be among the most promising
discovery channels at the LHC. They can provide sensitivity to dark matter, one of
the few extensions of the Standard Model which are known to exist, even though
the exact nature of dark matter, its mass(es) and coupling constants, are so far
completely unknown. In this paper, we have used new data on mono-jet ($j + \met$)
and mono-photon ($\gamma + \met$) final states to constrain a large class of
dark matter models, namely those in which dark matter--quark or dark matter--gluon
interactions exist
and can be described in the framework of effective field theory.
(We have discussed the validity of effective field theory, and the modifications
to our limits in cases where it is not valid, in sections~\ref{sec:lightmediators}
and \ref{sec:invhiggs}, see figures~\ref{fig:light-mediator} and \ref{fig:invisibleHiggs}.)

Since events in which dark matter is produced have a harder $\met$ spectrum
than Standard Model background processes, it is
advantageous to use rather hard cuts on the jet or photon transverse momentum
and on the missing energy. We have confirmed this expectation by comparing the
sensitivity of mono-jet samples with different cuts
(figure~\ref{fig:Lambda-monojet}) finding a clear advantage for the so called \texttt{veryHighPT} analysis. Using this ATLAS mono-jet analysis we set strong limits on a variety of different types of dark matter couplings
(figure~\ref{fig:alllambdabounds}), in particular vector, axial vector,
$t$-channel mediated scalar interaction with quarks and interactions with
gluons.

These limits can be converted into constraints on the dark
matter--nucleon scattering cross section measured in direct detection
experiments (figure~\ref{fig:dd-monojet}) and the dark matter annihilation
cross section (figure~\ref{fig:annihilation}).  For small dark matter mass,
$m_\chi \lesssim 5$~GeV, the LHC provides the strongest constraints for all
considered operators. At higher masses, direct detection experiments still have
an advantage if dark matter--nucleon scattering is spin-independent.  If dark matter couples primarily to gluons (for instance through a heavy quark
loop), the advantage is only marginal up to $m_\chi \sim 1$~TeV, where LHC
constraints deteriorate rapidly due to the limited center of mass energy.
For spin-dependent dark matter--nucleon scattering, the LHC constraints
surpass direct detection bounds by several orders of magnitude for dark
matter masses below the kinematic limit of the LHC.  
It should be noted that the collider constraints do not suffer from any astrophysical uncertainties, such as the (unknown) abundance of DM in the Earth's vicinity, or its velocity distribution.
Finally, we emphasize that if the DM--Standard Model coupling involves a light mediator, as discussed in section~\ref{sec:lightmediators}, the collider bounds may become considerably weakened.  If a direct detection experiment, spin-independent or spin-dependent, were to see an excess in apparent contradiction with these collider bounds, their existence would allow us to infer the presence of a light mediator---a fact we would be unaware of without these collider constraints.

As far as
limits on dark matter annihilation are concerned, the LHC is able to rule
out dark matter with thermal relic cross sections for $m_\chi \lesssim 15$~GeV
for vector couplings to quarks, and for $m_\chi \lesssim 70$~GeV for
axial vector couplings to quarks.
Limits from the mono-photon channel (figures~\ref{fig:alllambdaboundsphoton}
and \ref{fig:dd-monophoton}) are somewhat weaker than those from the
mono-jet channel (figures~\ref{fig:alllambdabounds} and \ref{fig:dd-monojet}), but not by much.  Furthermore, since they probe a different set of operators and suffer from different systematic uncertainties they provide a useful complementary search channel giving insight into the couplings of DM should an excess be found in either channel.

In the final section of this paper, we have considered a more specific type of
dark matter, interacting through a ``light mediator'', namely the Standard Model
Higgs boson $h$. If the decay channel $h \to \bar\chi \chi$ is kinematically
allowed, we have found that the most stringent constraints on dark matter
interactions can be derived from searches for invisible Higgs decays in the
$Z + H$ and vector boson fusion (VBF) production channels. Amusingly, for
certain Higgs mass ranges, it is possible in this framework to also set
\emph{lower} limits on dark matter--Standard Model interactions. In particular,
if the Higgs boson has a mass that is already excluded within the Standard Model,
the model can be reconciled with the data if the Higgs branching fraction
into dark matter is sufficiently large, which limits the dark matter--Higgs
couplings from \emph{below}. This lower bound on direct detection is already in conflict with bounds from XENON-100 for some regions of parameter space. Within the Higgs-coupled DM framework, there is an interesting interplay between dark matter searches and SM Higgs boson searches at the LHC. This interplay can be interesting and non-trivial, both in the case of new bounds on the Higgs and in the case of a SM Higgs discovery.

The analyses in this paper were carried out for the 7 TeV LHC on an integrated luminosity of at most 1.14 fb$^{-1}$, a tiny fraction of what we hope to accumulate in the coming years.  The increased statistics, and higher center of mass energy, will improve not only the ability to harden the cuts, making the analyses more sensitive to DM, but also bring the systematic uncertainties under greater control.  With dedicated analyses from both LHC collaborations, as well as searches on the final Tevatron dataset, we can expect great improvements on the bounds, or perhaps even the first observation of production of DM in the lab.

\section*{Acknowledgments}
We thank David Berge, Yuri Gershtein, Gordan Krnjaic, Joe Lykken, Adam Martin, Tom Schwarz, Shalhout Shalhout, Tim Tait and Steven Worm for helpful discussions.  PF, RH and YT thank the KITP Santa Barbara for hospitality during completion of part of this work. JK is grateful to the Aspen Center for Physics (supported by the National Science Foundation under Grant No.~1066293) for hospitality during part of this work. 
During this work YT was supported by a Fermilab Fellowship in Theoretical Physics.
Fermilab is operated by Fermi Research Alliance, LLC, under Contract DE-AC02-07CH11359 with the United States Department of Energy.

\appendix
\section{Invisible Higgs Decays in View of LHC Data (added in June 2014)}
\label{sec:lhc-data}

\begin{figure}
  \begin{center}
    \includegraphics[height=0.45\textwidth]{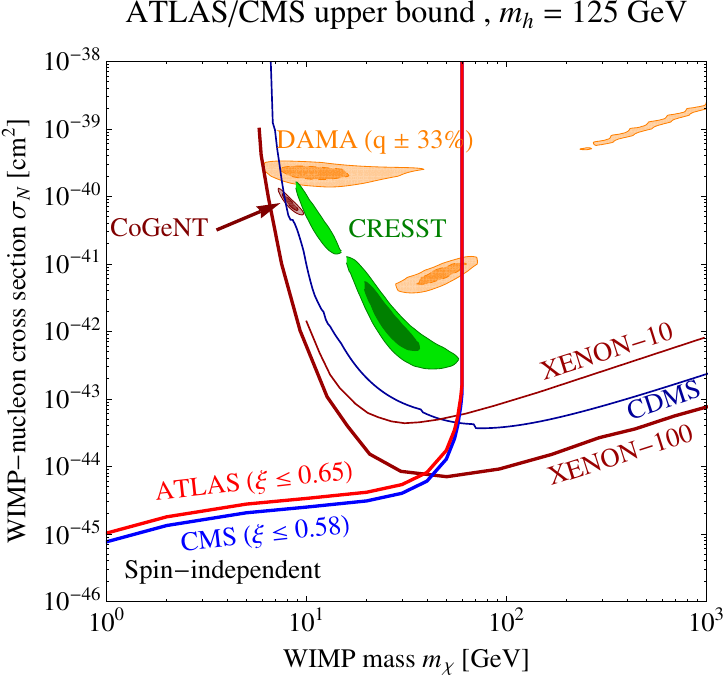}
  \end{center}
  \caption{Projected 95\% C.L.\ upper bounds on dark matter--nucleon scattering
    mediated by a Higgs boson from searches for invisible Higgs decays in
    ATLAS~\cite{ATLAS-CONF-2013-011} and
    CMS~\cite{Chatrchyan:2014tja} with combined 7~TeV and 8~TeV data. Limits are
    set by the $Z+H$ and vector boson fusion (VBF) production modes.}
  \label{fig:invisibleHiggs-newdata}
\end{figure}

Here, we update figure~\ref{fig:invisibleHiggs}, taking into account the
results from Run~1 of the LHC. Using the full 7~TeV and 8~TeV data set and
studying the $ZH$ and VBF channels, the ATLAS~\cite{ATLAS-CONF-2013-011} and
CMS~\cite{Chatrchyan:2014tja} collaborations have constrained the invisible
branching ratio $\xi$ of the Higgs boson to be $\xi < 65\%$ (ATLAS) and $\xi < 58\%$
(CMS), respectively, at 95\% C.L. In figure~\ref{fig:invisibleHiggs-newdata},
we translate these limits into constraints on the Higgs mediated DM--nucleon
scattering cross section.

\bibliographystyle{apsrev}
\bibliography{lhc-dm}

\end{document}